\documentclass[epjCONF, onecolumn]{svjour}
\usepackage{graphics}
\usepackage{graphicx}
\usepackage{bm}
\usepackage[varg]{txfonts}
\usepackage[latin1]{inputenc}
\session-title{Hadron Nuclear Physics (HNP) 2011}

\newcommand{\be}{\begin{equation}}
\newcommand{\ee}{\end{equation}\noindent}
\newcommand{\bea}{\begin{eqnarray}}
\newcommand{\eea}{\end{eqnarray}}

\newcommand{\maprightb}[1]{\smash{\mathop{
\hbox to 1cm{\rightarrowfill}}\limits_{#1}}}

\newcommand{\bc}{\begin{center}}
\newcommand{\ec}{\end{center}}

\newcommand{\matTwo}{\left(\begin{array}{rr}}
\newcommand{\matThree}{\left(\begin{array}{rrr}}
\newcommand{\emat}{\end{array}\right )}
\newcommand{\detTwo}{\left|\begin{array}{rr}}
\newcommand{\detThree}{\left|\begin{array}{rrr}}
\newcommand{\edet}{\end{array}\right |}

\begin{document}
\title{QCD Phase Diagram with Imaginary Chemical Potential}
\author{Keitaro Nagata \thanks{\email{kngt@hiroshima-u.ac.jp}}
and  Atsushi Nakamura}
\institute{Research Institute for Information Science and Education,
Hiroshima University, Higashi-Hiroshima 739-8527 Japan}
\abstract{We report our recent results on the QCD phase diagram obtained from 
the lattice QCD simulation. The location of the phase boundary between hadronic 
and QGP phases in the two-flavor QCD phase diagram is investigated. 
The imaginary chemical potential approach is employed, which is based on 
Monte Carlo simulations of the QCD with imaginary chemical potential 
and analytic continuation to the real chemical potential region. 
}
\maketitle

\section{Introduction}

The QCD phase diagram, which illustrates states of matter formed in 
terms of the strong interaction at a temperature and chemical potential, 
has been of prime interest in recent 
physics covering particle physics, hadron/nuclear physics and astrophysics. 
On top of ordinary nuclear matter and hot or dense matter such as QGP or compact stars, 
a very rich structure has been predicted in the QCD phase diagram using  
many phenomenological studies. 
Thoroughgoing analyses of heavy ion data 
show that we are sweeping finite temperature and density
regions. See Ref.~\cite{Andronic:2009gj}.

First-principle calculations based on QCD are now highly called. 
If such calculations would be at our hand,
their outcomes are also very valuable for many research fields:
high energy heavy ion collisions, the high density interior of 
neutron stars and the last stages of the star evolution.
Needless to say, the inside of nucleus is also a baryon rich
environment, and lots of contributions to nuclear physics 
could be expected.

Because QCD is non-perturbative in most regions of the QCD phase diagram, 
one is forced to use the lattice QCD in order to obtain a quantitative 
understanding. The lattice QCD is expected to provide reliable information on 
the phase structure based on QCD. Indeed, recently, there have been many active 
quantitative investigations about the finite temperature QCD~\cite{Fodor:2010,DeTar:2011nm}. 

On the other hand, simulations of systems with non-zero quark chemical 
potential $\mu$ have been a long challenge for the lattice QCD because 
of the notorious sign problem. 
In the lattice QCD, a fermionic determinant $\det \Delta(\mu)$ is used as 
a probability in a Monte Carlo method. The introduction of non-zero $\mu$ 
makes $\det \Delta(\mu)$ complex, and therefore leads to the breakdown of the 
stochastic part of the lattice QCD.
Despite of the severe sign problem, several approaches have been 
proposed to study the QCD with nonzero $\mu$, where the location of the 
phase boundary and critical endpoint, and the determination of EoS have been 
extensively investigated. See e.g.~\cite{Muroya:2003qs,deForcrand:2010ys}. 

One idea is to perform simulations in systems with an imaginary 
chemical potential, where the sign problem is absent. 
The phase diagram in the imaginary chemical potential region is connected to 
that in the real chemical potential region, i.e., ordinary QCD phase diagram, 
owing to the analytic continuation. 
In addition, HMC algorithms are available without any truncation for the quark 
determinant containing imaginary chemical potential. 
Numerical costs are relatively small compared to approaches 
requiring the direct evaluation of the quark determinant.
Hence, the lattice QCD with imaginary chemical potential is one of the 
standard technique and has been studied by using staggered fermions 
with two flavor~\cite{deForcrand:2002ci,D'Elia:2009qz,D'Elia:2009tm}, 
three flavor~\cite{deForcrand:2010he}, four flavor~\cite{D'Elia:2002gd,D'Elia:2004at,D'Elia:2007ke,Cea:2010md} 
in 2-color QCD and finite isospin QCD~\cite{Cea:2009ba,Cea:2007vt}, 
Wilson fermions with two flavor~\cite{Wu:2006su} and clover-improved 
Wilson fermions with two flavor~\cite{Nagata:2011yf}. In addition, 
data obtained in such a simulation are also useful the matching of 
phenomenological models such as Polyakov loop extended 
Nambu-Jona-Lasinio(PNJL) models with the lattice QCD~\cite{Sakai:2008py,Kouno:2009bm,Sasaki:2011wu,Sakai:2011gs}.

Recently, we have studied the two-flavor QCD phase diagram using the lattice QCD simulations~\cite{Nagata:2010xi,Nagata:2011yf}. 
We have used the imaginary chemical potential approach to avoid the sign problem. 
Here, we report our results on the study of the QCD phase diagram. 

\section{QCD with Imaginary Chemical Potential}
\label{Ap2111sec1}

Let chemical potential complex $\mu= \mu_R + i\mu_I (\mu\in\mathbb{C},\; \mu_R, \mu_I\in\mathbb{R})$. 
Fermion determinants satisfy a relation 
\begin{eqnarray}
\Delta(\mu)^\dagger = \gamma_5 \Delta( - \mu^*) \gamma_5,
\label{Jan1111eq1}
\end{eqnarray}
where $\Delta(\mu)$ is a quark matrix. It is straightforward from Eq.~(\ref{Jan1111eq1}) to obtain 
$(\det \Delta(\mu) )^* = \det\Delta( - \mu^*)$.
This implies that $\det \Delta(\mu)$ is complex for a real chemical potential 
$\mu=\mu_R$, which causes the sign problem. 
On the other hand, one can easily prove $\det \Delta(\mu)$ is real 
for a pure imaginary chemical potential $\mu= i\mu_I$.
The sign problem does not occur in this case, and Monte Carlo methods 
are available. 

\begin{figure}[htbp]
\begin{center}
\includegraphics[height=.20\textheight]{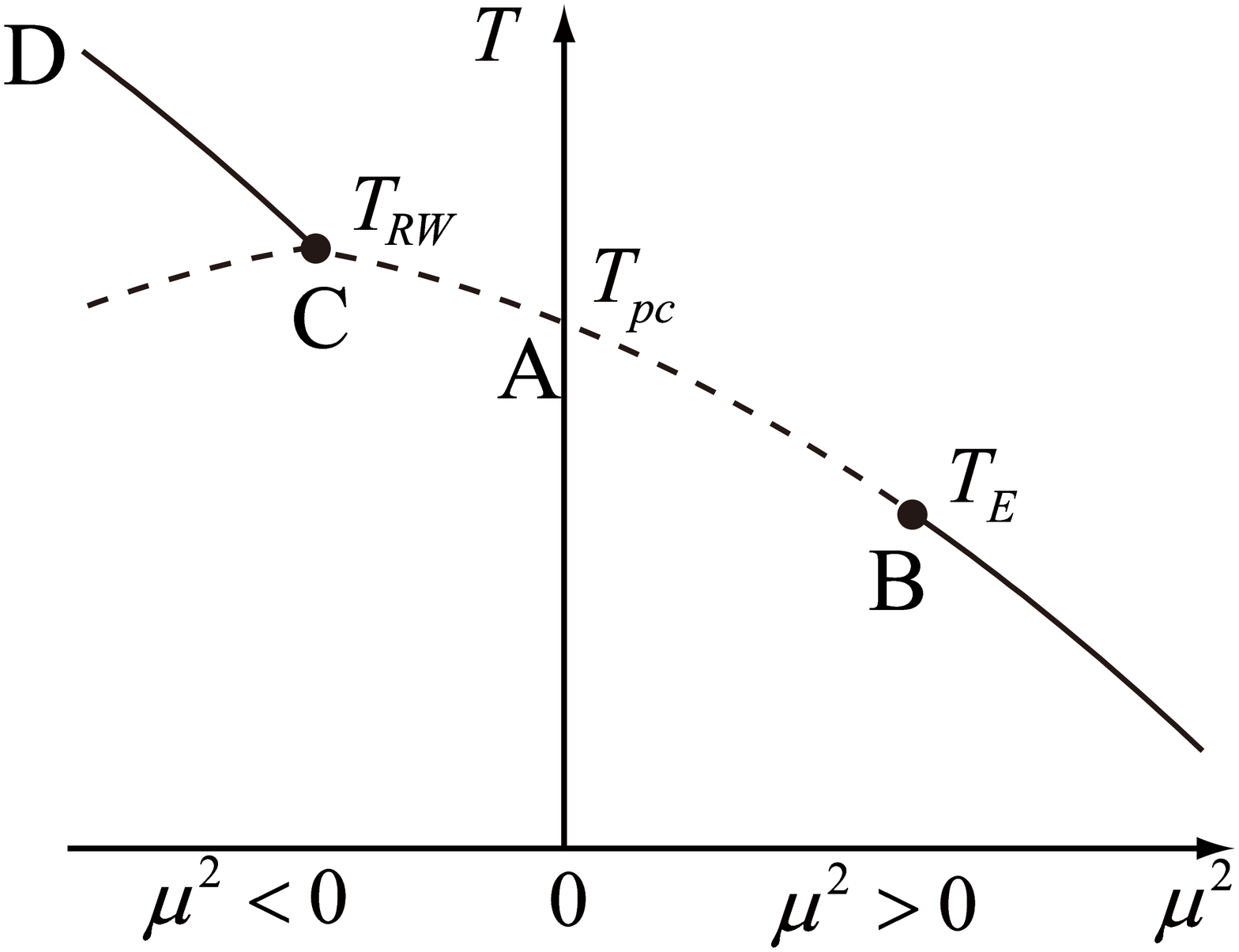}
\includegraphics[height=.20\textheight]{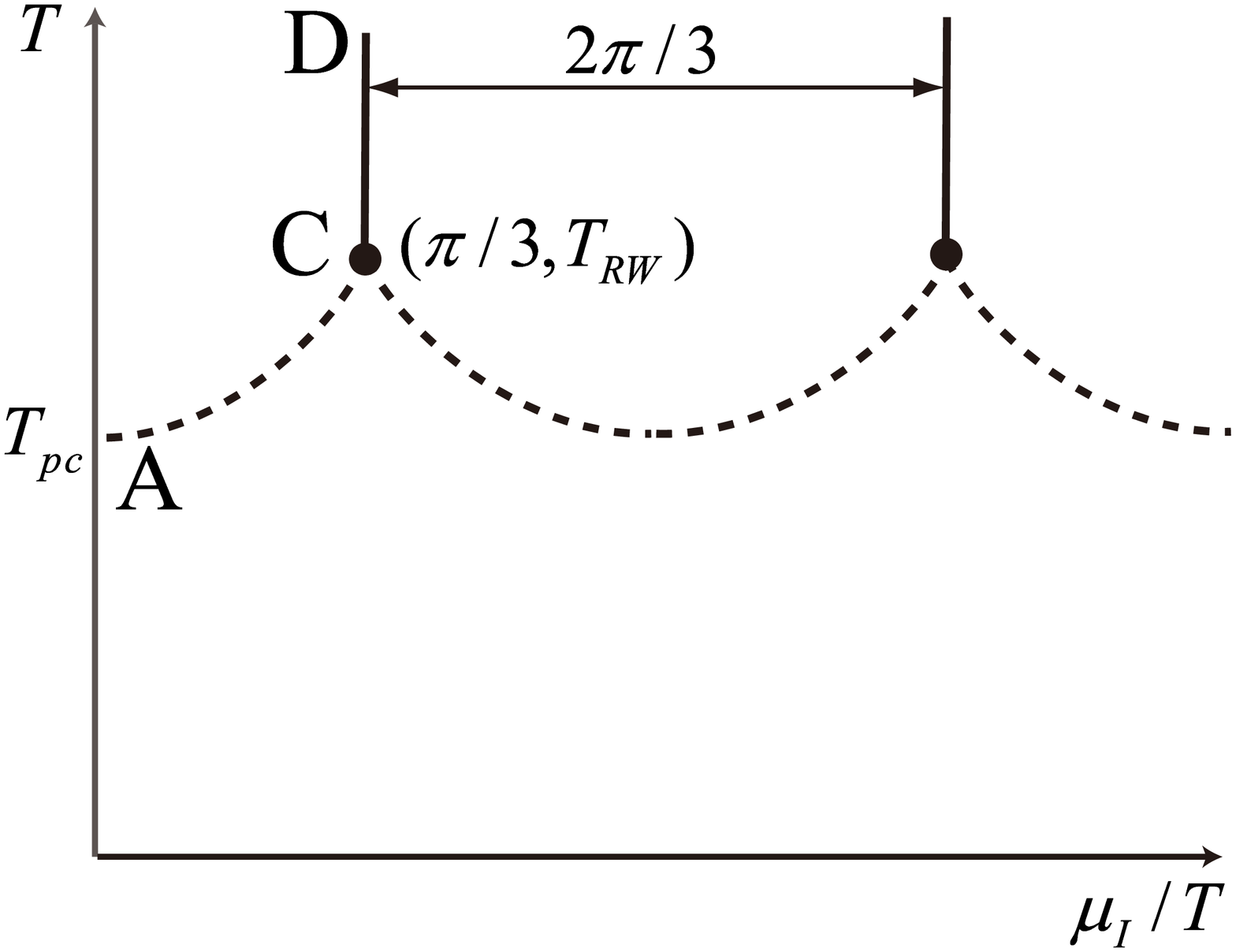}
\begin{minipage}{12cm}
\caption{Schematic figures for the $N_f=2$ QCD phase diagram in 
the $(\mu^2, T)$ plane (left) and $(\mu_I/T, T)$ plane (right). 
A : Pseudo-critical point at $\mu=0$. B : Critical endpoint. 
C : Roberge-Weiss endpoint. 
AB : Pseudo-critical line. AC : Extension of the line AB into the 
imaginary chemical potential plane. CD : Roberge-Weiss phase 
transition line $\mu_I/T=\pi/3$. In the right panel, larger 
$\mu_I/T$ region of the phase diagram is obtained from  the RW periodicity.
}\label{Jan2311fig1}
\end{minipage}
\end{center}
\end{figure}%

Two characteristics of the $\mu^2\le 0$ region are so-called Roberge-Weiss(RW) 
phase transition and Roberge-Weiss(RW) periodicity~\cite{Roberge:1986mm}. 
The QCD grand partition function has a periodicity with a period $2\pi/N_c$ as 
\begin{eqnarray} 
Z\left(\frac{\mu_I}{T}\right)= Z\left(\frac{\mu_I}{T} + \frac{2\pi k }{N_c}\right), 
\end{eqnarray} 
where  $k$ is an integer. Furthermore, Roberge and Weiss showed from 
a perturbative analysis the existence of a first-order phase transition 
on the line $\mu_I/T=\pi/N_c$, and from a strong coupling analysis 
the absence of such a transition at low temperatures. 
Because the RW phase transition occurs at high temperatures but does not at low temperatures, 
it may have an endpoint at a temperature $T_{RW}$ on the line $\mu_I/T=\pi/3$. 
The order of the point has been extensively investigated in Ref.~\cite{D'Elia:2009qz,deForcrand:2010he} and 
the quark-mass dependence is found: first order for small and large quark masses 
and second order for an intermediate quark masses. 

The left panel of Fig.~\ref{Jan2311fig1} shows an expected 
phase diagram, where we employ the $(\mu^2, T)$ plane containing
both the real $(\mu^2\ge 0,  \mu=\mu_R)$ and imaginary $(\mu^2\le 0, \mu=i\mu_I)$ regions. 
Even if $\mu^2\le 0$, it is expected quark-gluon-plasma(QGP) 
and hadronic phases exist at high and low  temperatures, respectively. 
The two phases are separated by the pseudocritical line for the deconfinement crossover, 
which is an extension from the $\mu^2 \ge 0$ region. 
The absolute value of the Polyakov loop is often employed to identify 
confinement/deconfinement phase, although it is not a real order
parameter because of the crossover nature of the transition. 
The features in the $\mu^2\le 0$ region are well manifested in 
the $(\mu_I/T, T)$-phase diagram, see the right panel of Fig.~\ref{Jan2311fig1}. 

\section{Result}
\subsection{Set Up}
We employ a clover-improved Wilson fermion action of two-flavors and a 
renormalization-group improved gauge action. The clover-improved Wilson 
fermion action is given by 
\begin{eqnarray}
&\Delta(x,y)&  =   \delta_{x, x^\prime} 
 - \kappa \sum_{i=1}^{3} \left[
(1-\gamma_i) U_i(x) \delta_{x^\prime, x+\hat{i}} 
+ (1+\gamma_i) U_i^\dagger(x^\prime) \delta_{x^\prime, x-\hat{i}}\right] \nonumber \\
 &-&
\hspace{-4mm}
\kappa \left[ e^{+\mu} (1-\gamma_4) U_4(x) \delta_{x^\prime, x+\hat{4}}
+e^{-\mu} (1+\gamma_4) U^\dagger_4(x^\prime) \delta_{x^\prime, x-\hat{4}}\right] \nonumber 
 -  \kappa  C_{SW} \delta_{x, x^\prime}  \sum_{\mu \le \nu} \sigma_{\mu\nu} 
F_{\mu\nu}.
\label{April2111eq1}
\end{eqnarray}
Here $\mu$ is the quark chemical potential in lattice unit, which is introduced to 
the temporal part of link variables.

In order to scan the phase diagram, simulations were done 
for more than 150 points on the $(\mu_I, \beta)$ plane in the domain 
$0\le \mu_I \le  0.28800$ and $1.79 \le \beta \le 2.0$. 
All the simulations were performed on a $N_s^3\times N_t = 8^3\times 4$ lattice. 
The RW phase transition line in the present setup is given by 
$\mu_I = \pi/12\sim 0.2618$. 
The value of the hopping parameter $\kappa$ were determined for each value of $\beta$ 
according to a line of the constant physics with $m_{PS}/m_V=0.8$ obtained in 
Ref.~\cite{Ejiri:2009hq}. 

\begin{figure*}[htbp]
\begin{center}
\includegraphics[height=.21\textheight]{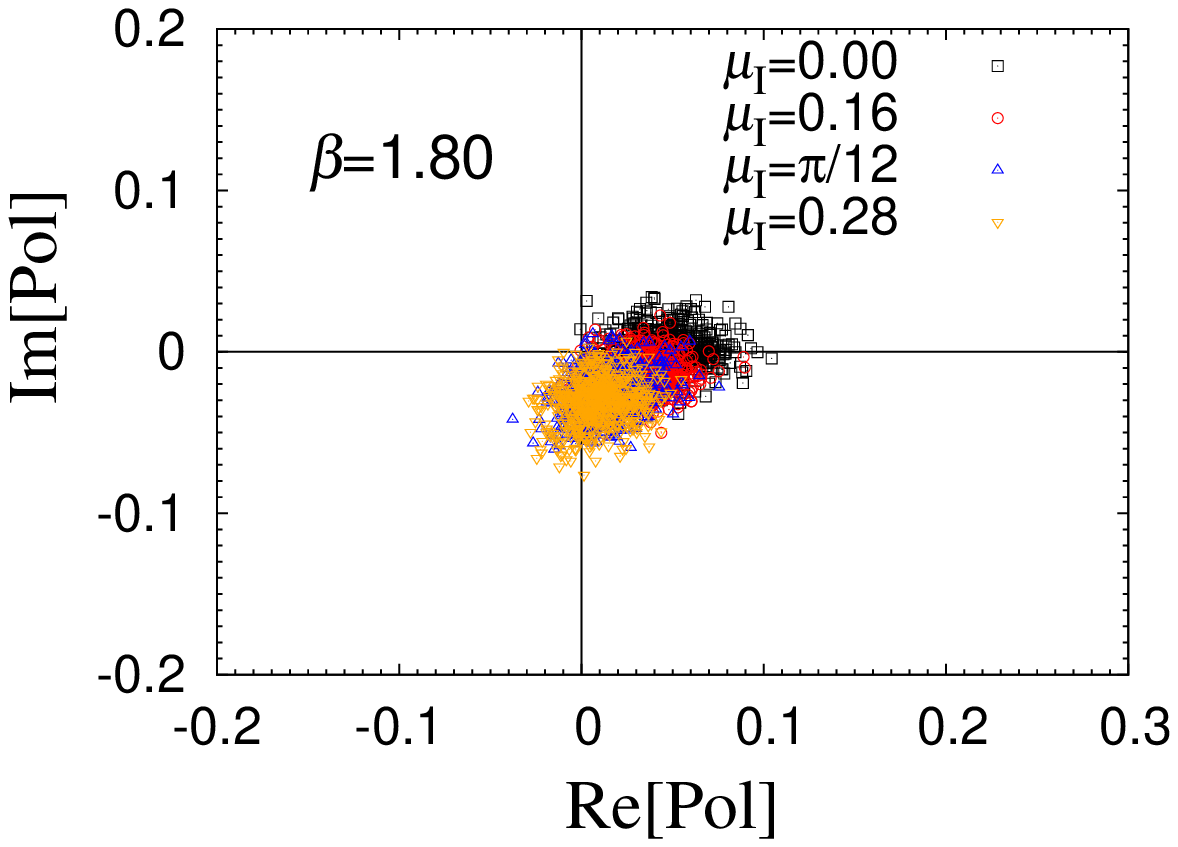}
\includegraphics[height=.21\textheight]{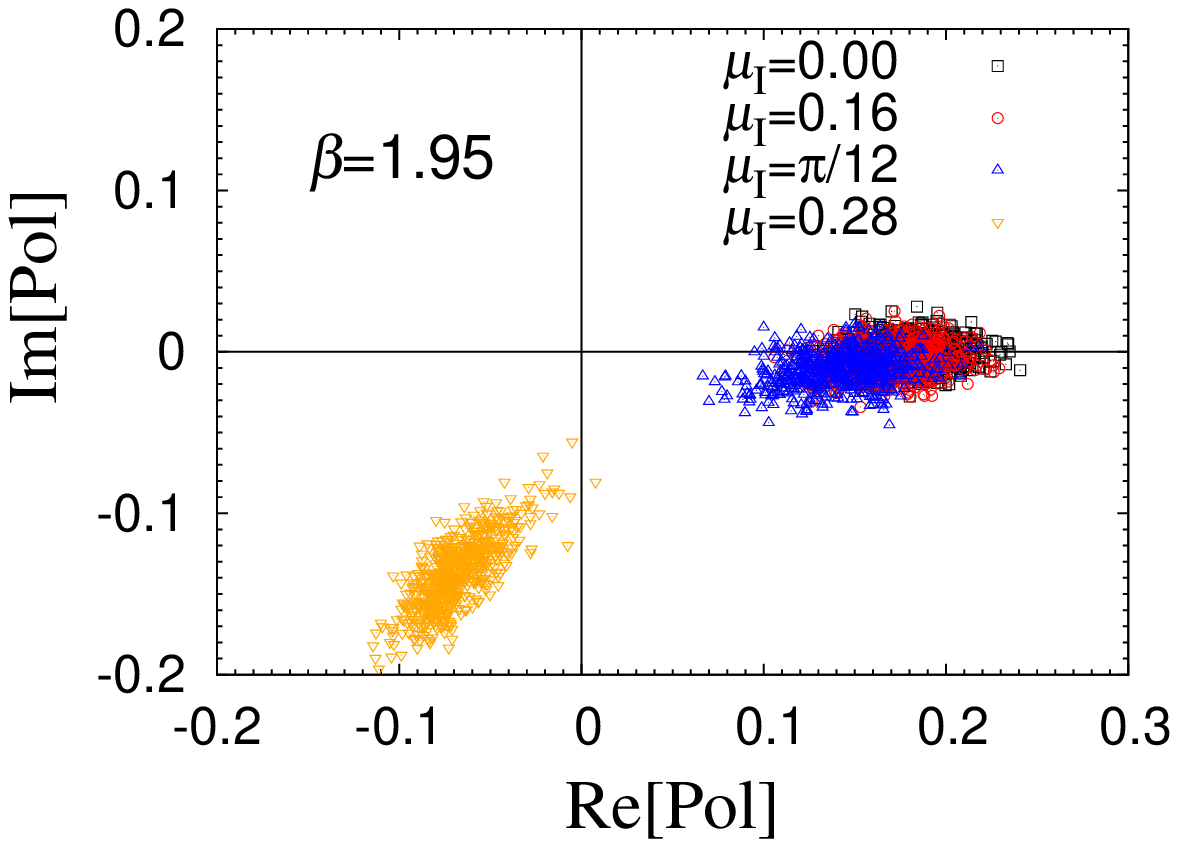}
\begin{minipage}{12cm}
\caption{Scatter plots of the Polyakov loop. Left : $\beta=1.80$ 
( $T<T_{pc}$).  Right : $\beta=1.95$  ( $T>T_{RW}$). 
}\label{Mar0411fig1}
\end{minipage}
\end{center}
\end{figure*}

Scatter plots of the Polyakov loop in the complex plane  are 
shown in Fig.~\ref{Mar0411fig1}, where we choose two typical cases 
$\beta=1.80$ for the hadronic phase and $\beta=1.95$ for the QGP phase.  
At low temperatures, the Polyakov loop is small in magnitude for any
$\mu_I$ and continuously changes in a clockwise direction
as increasing $\mu_I$. 
On the other hand, at high temperatures, the Polyakov loop grows to 
$0.2\sim 0.3$. It stays at the real axis for $\mu_I<\pi/12$ and 
jumps to the left-lower side at $\mu_I=\pi/12$. 
The difference of the Polyakov loop modulus between high and low 
temperatures shows the deconfinement crossover, which is the curve 
AC in Fig.~\ref{Jan2311fig1}. The observed jump of the Polyakov loop 
at $\mu_I=\pi/12$ is the Roberge-Weiss phase transition, which is the 
line CD. 

\subsection{Phase transitions and universality}

Now we discuss the properties of the phase transitions in the 
imaginary chemical potential region of the QCD phase diagram. 
\begin{figure*}[htbp]
\begin{center}
\includegraphics[width=0.45\linewidth]{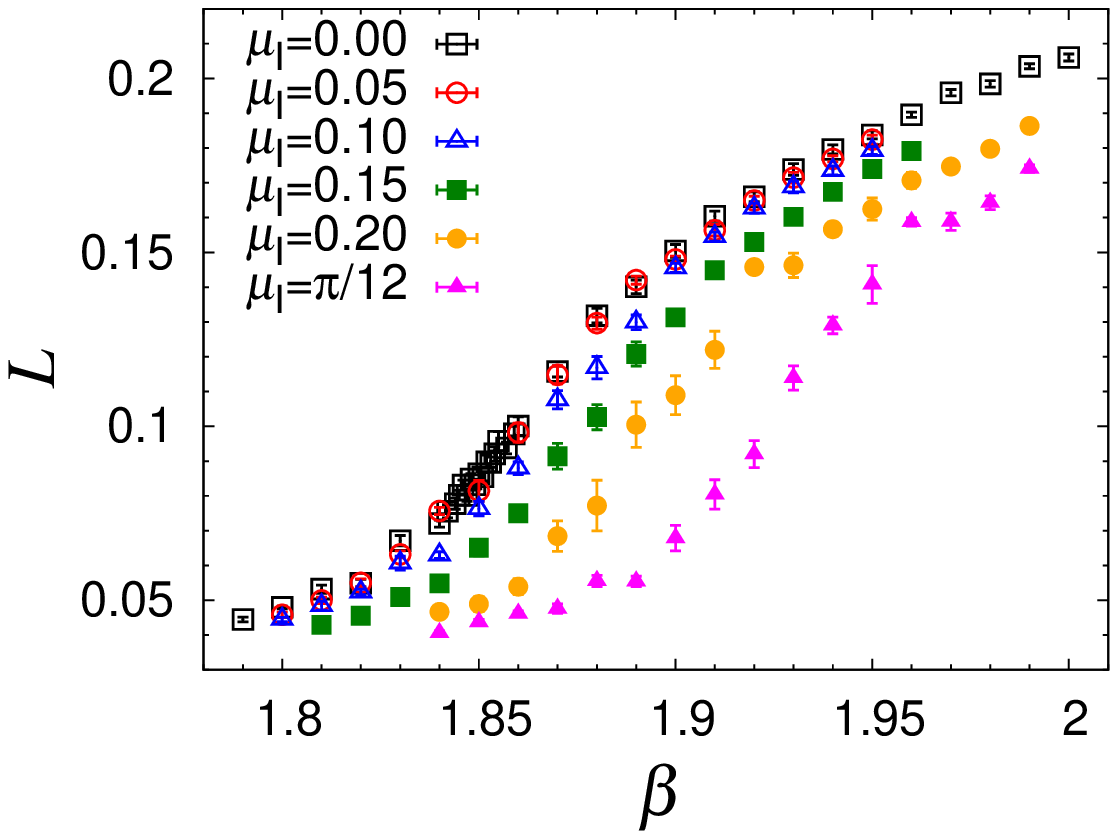}
\includegraphics[width=0.45\linewidth]{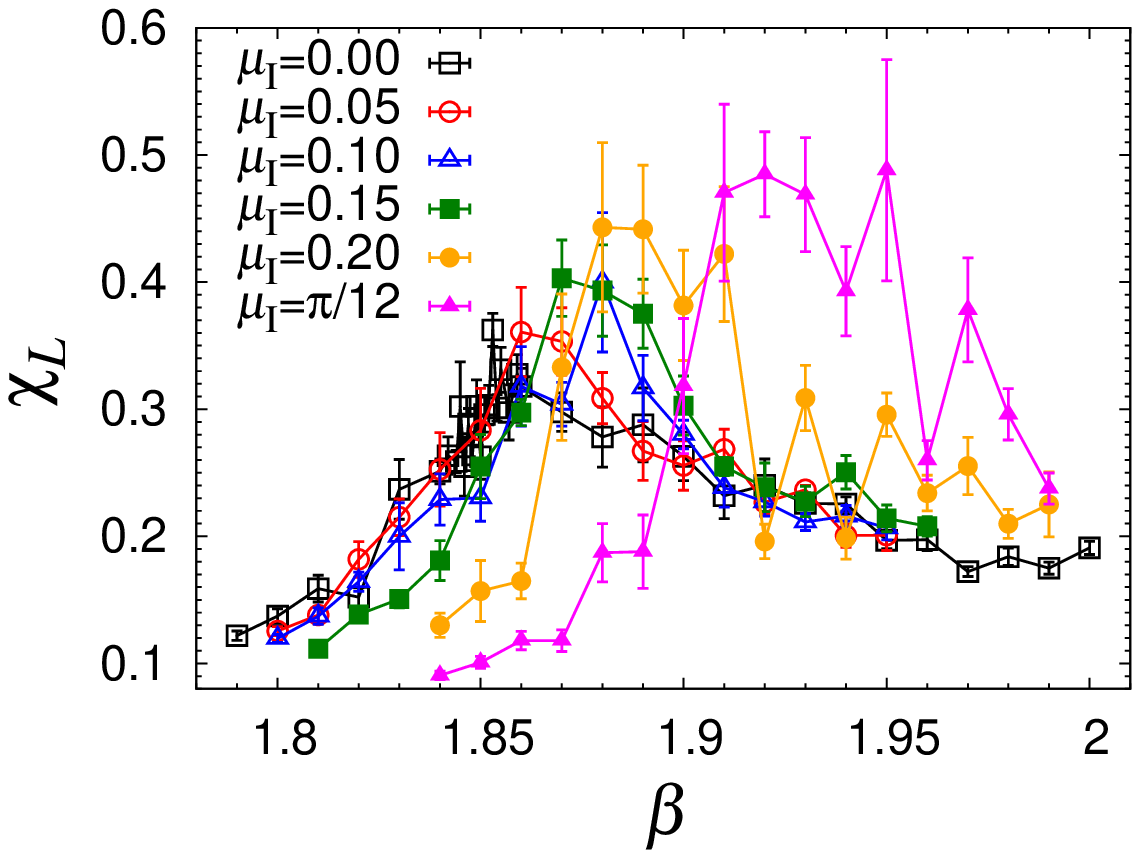}
\begin{minipage}{0.75\linewidth}
\caption{The $\beta$-dependence of the Polyakov loop modulus $L$ (left) and 
its susceptibility $\chi_L$ (right) for various $\mu_I$. 
}\label{Jan1011fig2}
\end{minipage}
\end{center}
\end{figure*}

The results of the Polyakov loop modulus $L$ and its susceptibility $\chi_L$ 
are shown in Fig.~\ref{Jan1011fig2}. 
Those behaviors  suggest the possibility that the system undergoes the 
crossover with increasing $\beta$ or temperature. 
It is understood from the $\mu_I$ dependence of the peak position of $\chi_L$ that 
the pseudo-critical temperatures become higher with increasing $\mu_I$ until $\mu_I=\pi/12$. 
This is consistent with the expected behavior shown in Fig.~\ref{Jan2311fig1}.

\begin{figure*}[htbp]
\begin{center}
\includegraphics[height=.35\textheight]{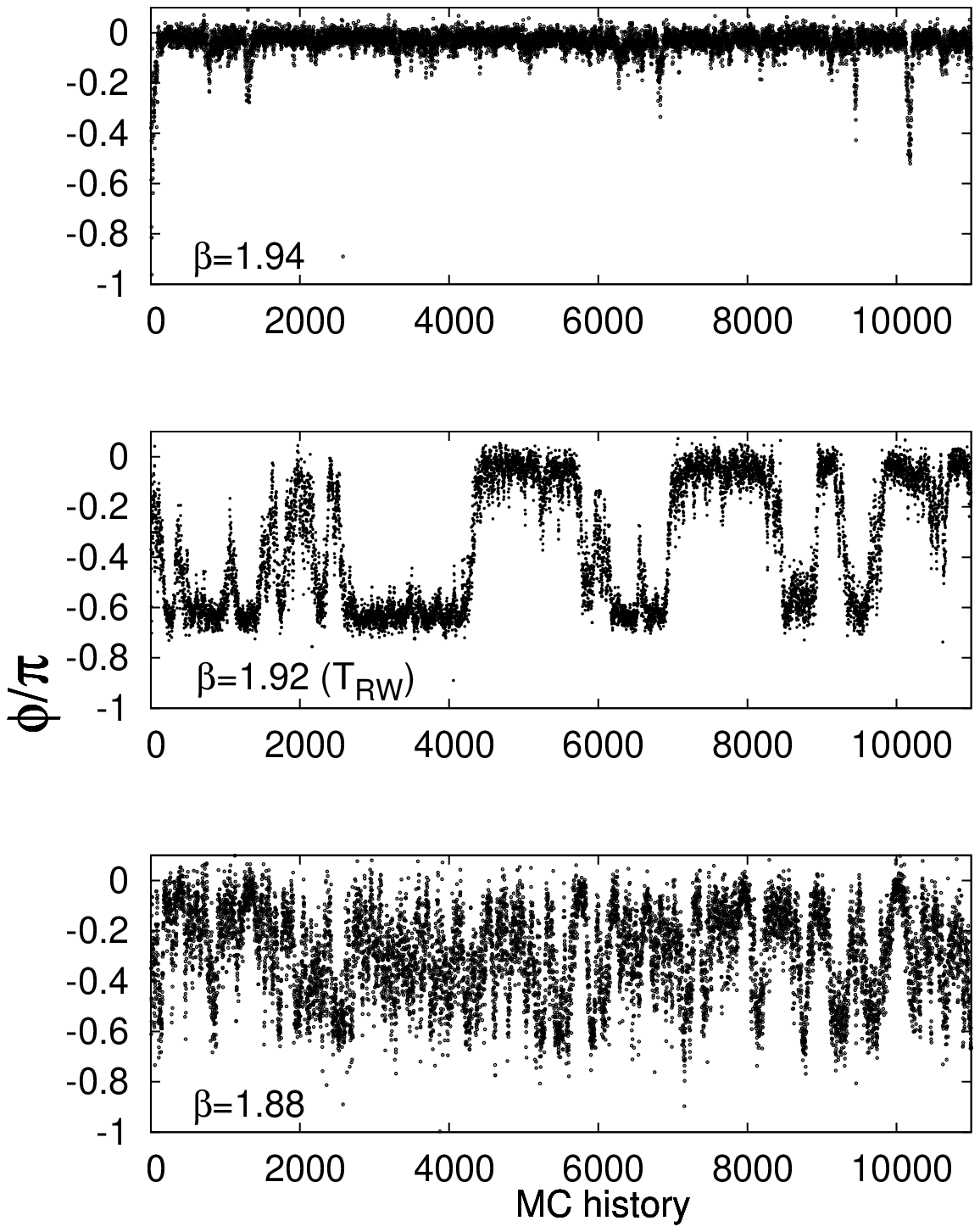}
\includegraphics[width=.4\linewidth]{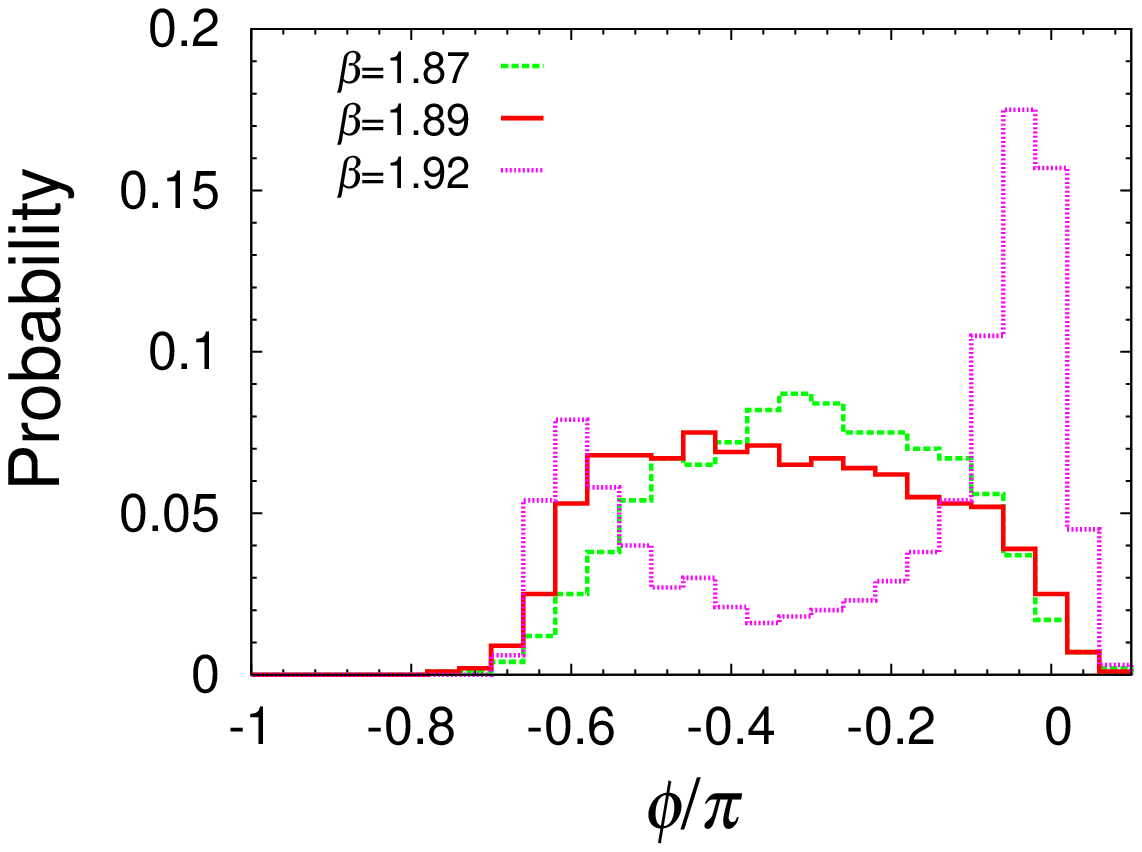}
\begin{minipage}{12cm}
\caption{
Monte Carlo history of the Polyakov loop phase at $\mu_I=\pi/12$ (left panels). 
The two state signal is found in the middle of the left panels.
The histogram of $\phi$ at $\mu_I=\pi/12$ for various $\beta$(right). 
}\label{Jun2311fig2}
\end{minipage}
\end{center}
\end{figure*}

At $\mu_I=\pi/12$, the system is on the RW phase transition line at high temperatures. 
Although the behavior of the $L$ and $\chi_L$ are similar, the two state 
signal is found for $\mu_I=\pi/12$, see Fig.~\ref{Jun2311fig2}.
\begin{figure*}[htbp]
\begin{center}
\includegraphics[height=.35\textheight]{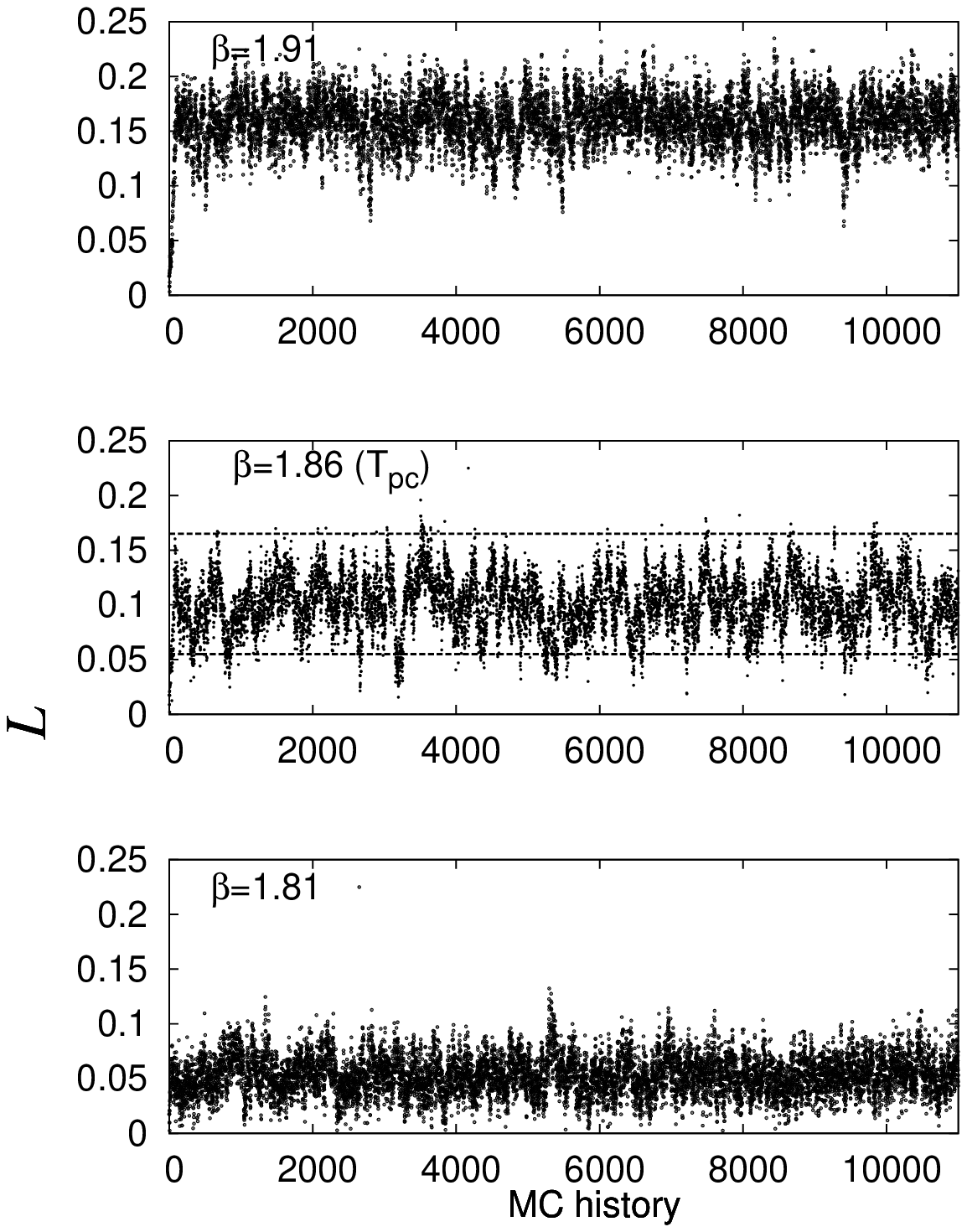}
\includegraphics[height=.35\textheight]{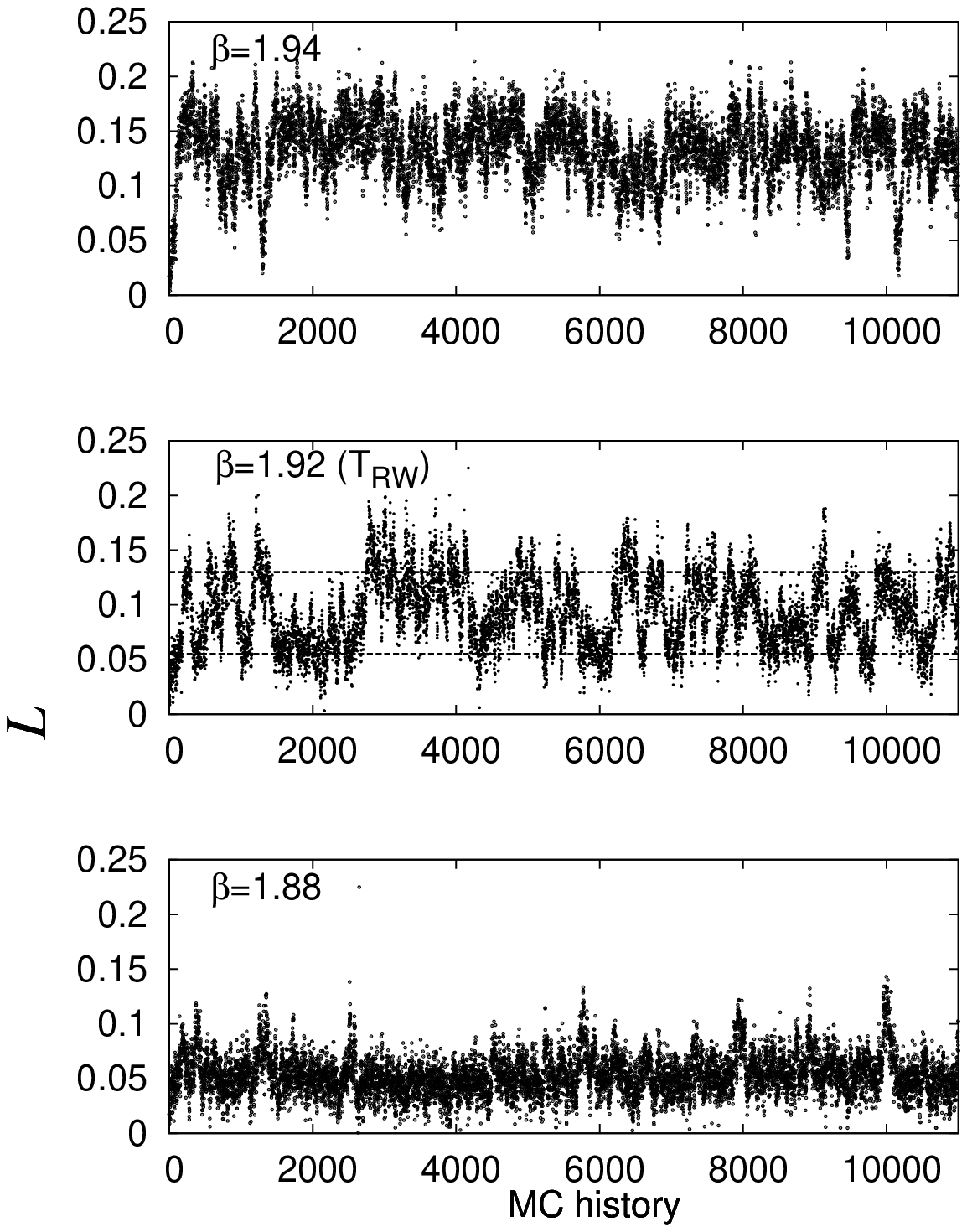}
\begin{minipage}{12cm}
\caption{Monte Carlo history of the Polyakov loop modulus at $\mu=0$ (left panels) and 
$\mu=\pi/12$ (right panels). Three  values of $\beta$ are chosen here. 
The two state signal is found in the middle of the right panels.
}\label{Jun2311fig1}
\end{minipage}
\end{center}
\end{figure*}

\begin{figure*}[htbp]
\begin{center}
\includegraphics[width=0.45\linewidth]{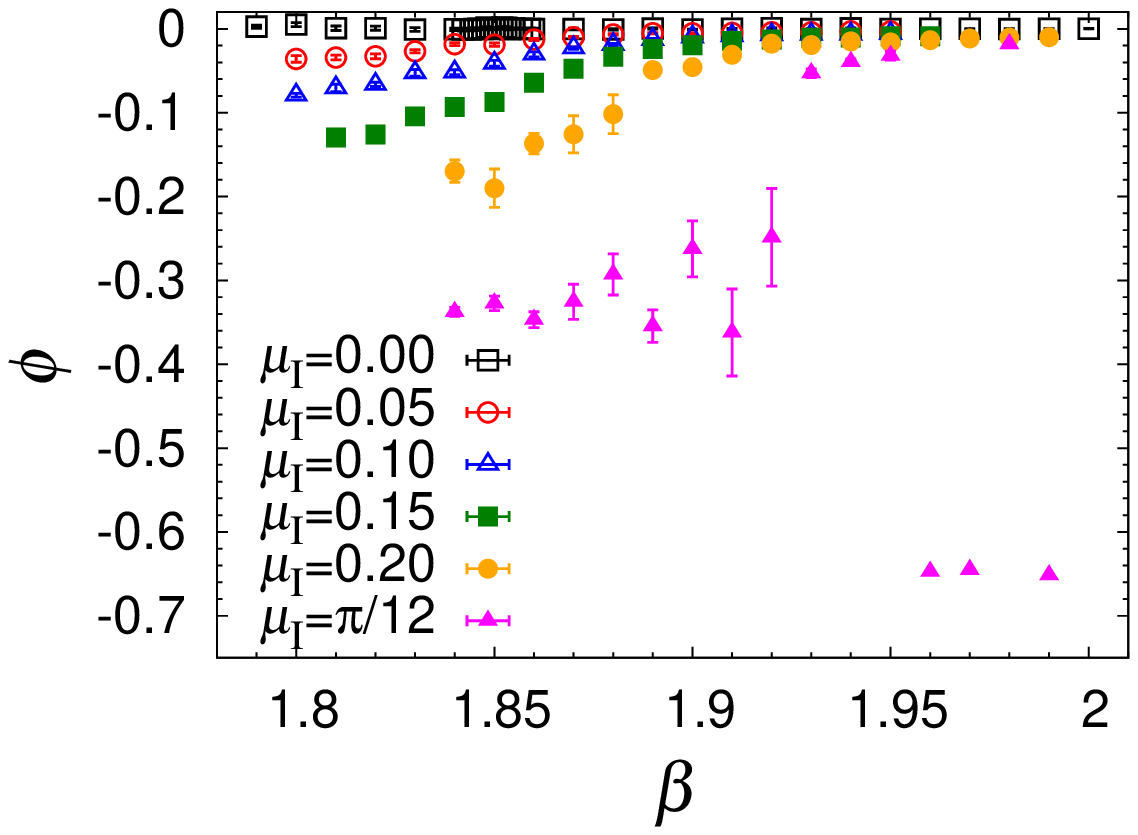}
\includegraphics[width=0.45\linewidth]{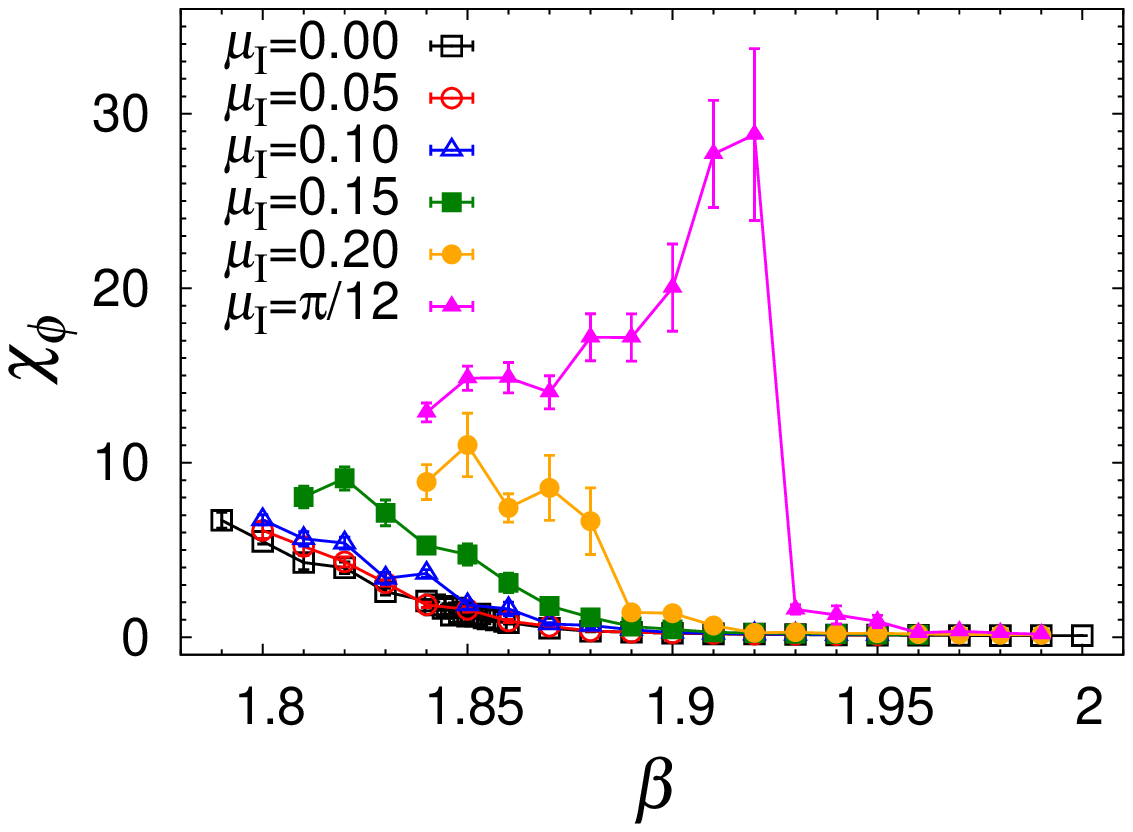}
\begin{minipage}{0.75\linewidth}
\caption{The $\beta$-dependence of the Polyakov loop phase $\phi$ (left) and 
its susceptibility $\chi_\phi$ (right) for various $\mu_I$.
}\label{Jan1011fig3}
\end{minipage}
\end{center}
\end{figure*}

It is more convenient to consider the Polyakov loop phase  $\phi$ and its susceptibility $\chi_\phi$ 
in order to study the nature of the RW endpoint, which are shown in Fig.~\ref{Jan1011fig3}. 
For $\mu_I=\pi/12$, $\phi$ rapidly changes near $\beta=1.92$.
It is seen  that for $\mu_I=\pi/12$ 
there is one vacuum at low temperatures and are two vacua at high temperatures. 
The histogram and Monte Carlo history of $\phi$ at $\mu_I=\pi/12$ in Fig.~\ref{Jun2311fig2} also 
show this behavior. 
The susceptibility $\chi_\phi$ shows a divergent-like behavior near 
$\beta=1.92$ only for $\mu_I=\pi/12$. 
These behaviors suggest the possibility that the system undergoes the second order 
phase transition at the RW endpoint with increasing temperature.  
We are now studying a finite size scaling analysis using lattice sizes $N_s=6, 10$. 
Preliminary result supports that the RW endpoint is second order. 

\begin{figure*}[htbp]
\begin{center}
\includegraphics[width=0.45\linewidth]{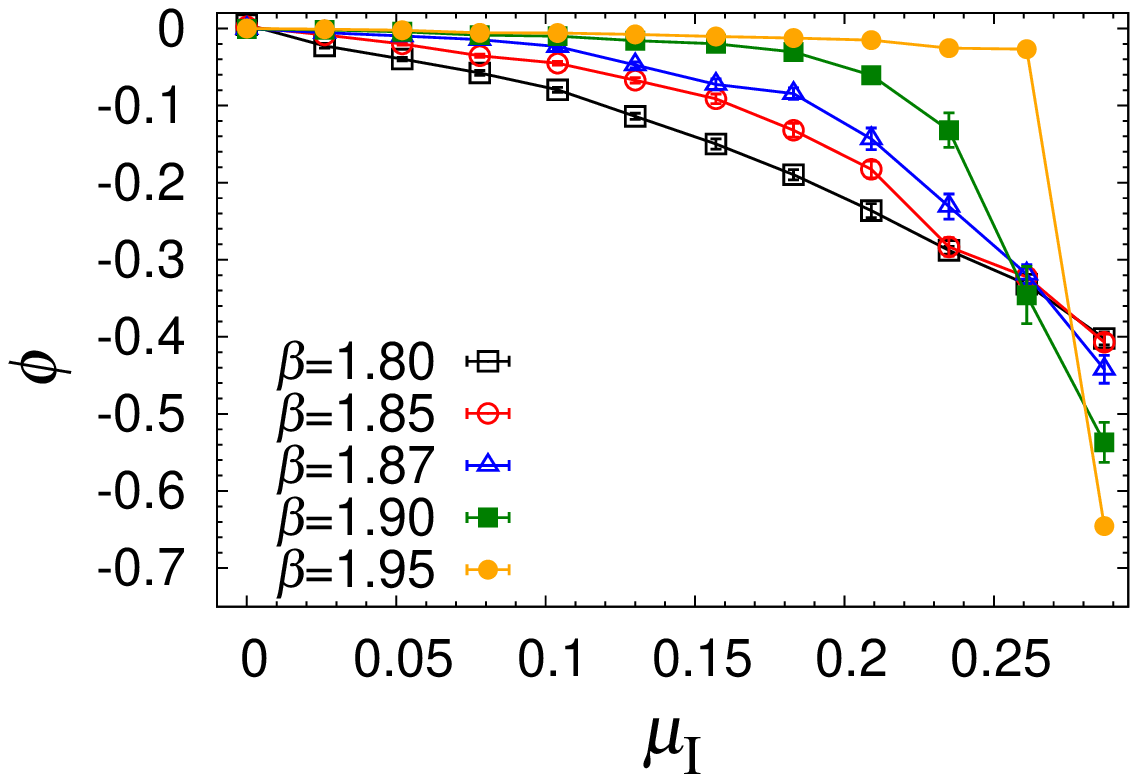}
\includegraphics[width=0.45\linewidth]{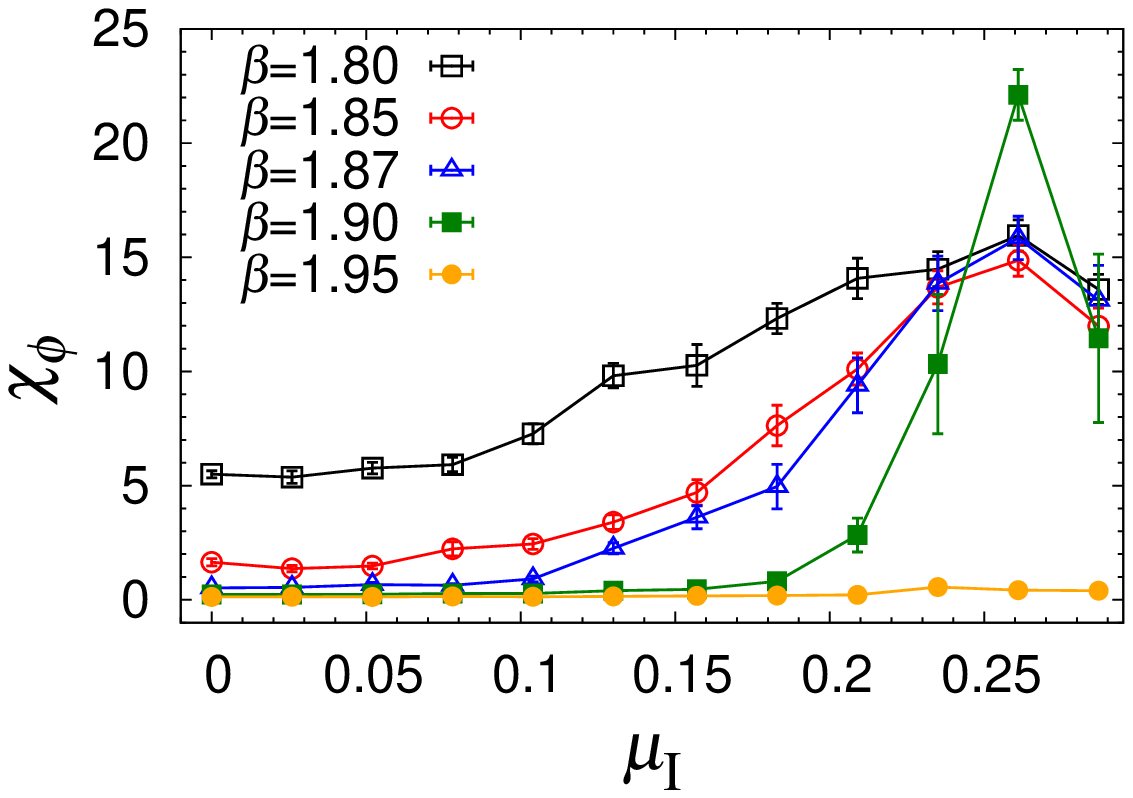}
\begin{minipage}{0.75\linewidth}
\caption{The $\mu_I$-dependence of $\phi$ and $\chi_\phi$ for various $\beta$.
}\label{Jan2111fig5}
\end{minipage}
\end{center}
\end{figure*}
The $\mu_I$-dependence of $\phi$ and $\chi_\phi$ are shown in Fig.~\ref{Jan2111fig5}. 
$\phi$ is a smooth function of $\mu_I$ at low temperatures ($\beta=1.80$-$1.90$), 
while $\phi$ jumps to $-2\pi/3$ from $0$ at $\mu_I=\pi/12$ at a high temperature 
($\beta=1.95)$. The system undergoes the first order phase transition at $\mu_I=\pi/12$ 
at high temperatures. Note $L$ is periodic and $\phi$ is anti-periodic, which is caused by 
the periodicity of the $\mu_I$-dependence of the Polyakov loop~\cite{Kouno:2009bm}.

\begin{figure*}[htbp]
\begin{center}
\includegraphics[height=.3\textheight]{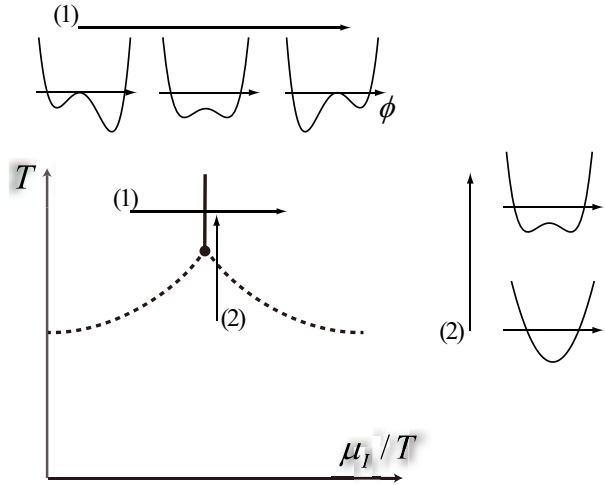}
\begin{minipage}{12cm}
\caption{Nature of the imaginary chemical potential region of the two-flavor 
QCD phase diagram with an intermediate quark mass obtained from the lattice 
QCD simulation with the clover-improved Wilson fermion.
}\label{June2811fig1}
\end{minipage}
\end{center}
\end{figure*}

Now, the properties of the QCD phase diagram with imaginary chemical potential is 
summarized in Fig.~\ref{June2811fig1}.

\subsection{Pseudocritical Line}

Next, we determine the pseudocritical line. 
Critical values of $\beta$ for the deconfinement crossover are 
obtained from the susceptibility of the Polyakov loop modulus for each $\mu_I$. 
We use the result obtained by WHOT collaboration~\cite{Ejiri:2009hq} 
to translate $\beta$ to $T$. 

\begin{figure*}[htbp]
\begin{center}
\includegraphics[height=.22\textheight]{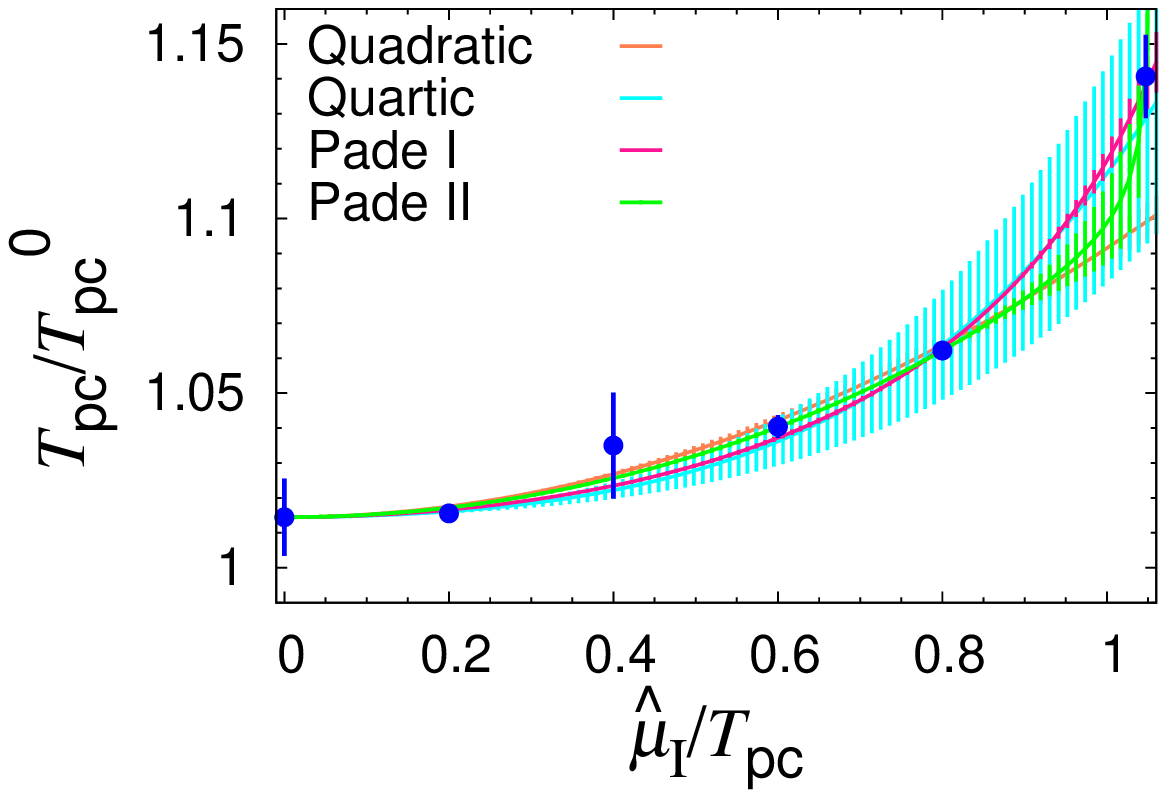}
\includegraphics[height=.22\textheight]{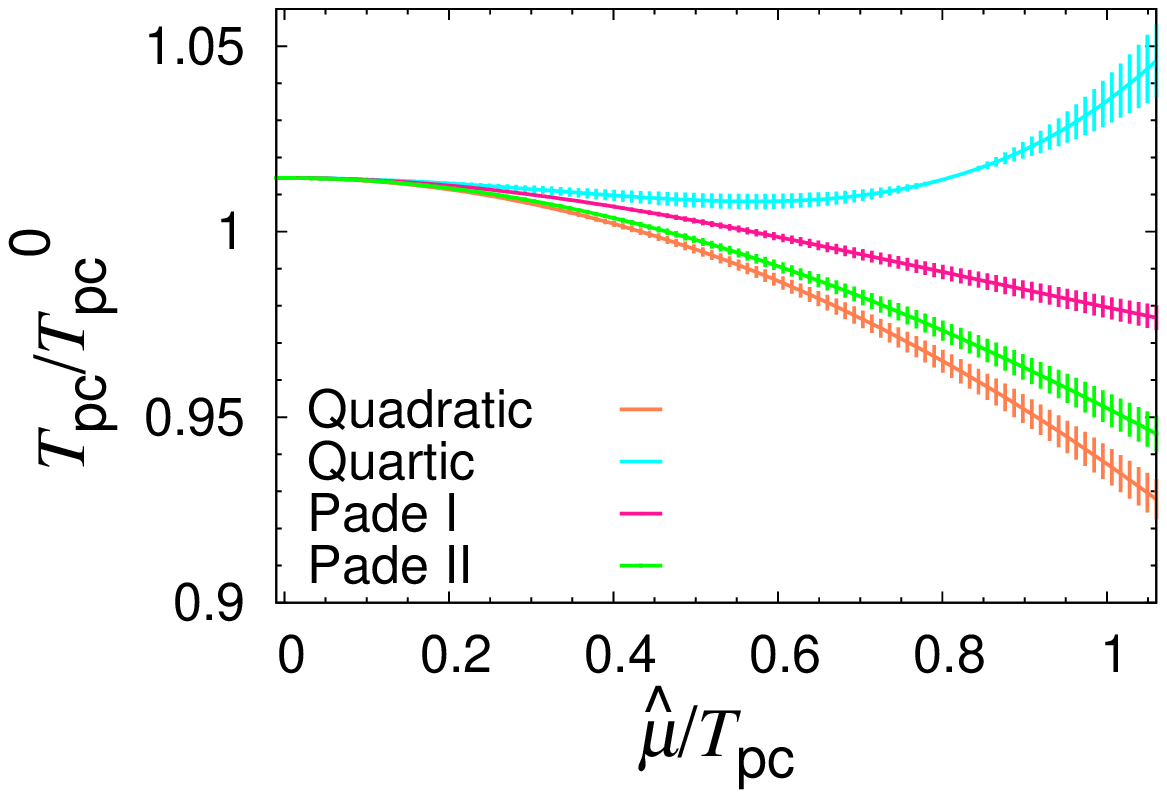}
\begin{minipage}{12cm}
\caption{The pseudo-critical line $\beta_{pc}$ in the imaginary 
(left panel) and real(right panel) region. 
}\label{Feb2711fig2}
\end{minipage}
\end{center}
\end{figure*}

We consider quadratic, quartic functions and two types of the Pad\'e approximation~\cite{Cea:2007vt,Cea:2009ba,Cea:2010md}; 
\begin{eqnarray}
\frac{T_{pc}}{T_{pc}^0} &=& \sum_n d_n \left(\frac{\hat{\mu}_I}{T_{pc}}\right)^{2n}, 
\label{Mar0611eq3} \\
\frac{T_{pc}}{T_{pc}^0} &=&  d_0 \frac{ 1 + d_1 (\hat{\mu}_I/T_{pc})^2}{1 + d_2 (\hat{\mu}_I/T_{pc})^2}\;\;\; \mbox{(Pad\'e (I))},
\label{Mar0611eq4}  \\
\left(\frac{T_{pc}}{T_{pc}^0}\right)^2 &=&  d_0 \frac{ 1 + d_1 (\hat{\mu}_I/T_{pc})^2}{1 + d_2 (\hat{\mu}_I/T_{pc})^2 + d_3  (\hat{\mu}_I/T_{pc})^4 }
\label{May0711eq1}   \;\;\; \mbox{(Pad\'e (II))} \nonumber, 
\end{eqnarray}
where $\mu_I = a \hat{\mu}_I$, and $\hat{\mu}_I$ is the imaginary chemical potential in 
physical unit. $T_{pc}^0$ and $T_{pc}$ are pseudo-critical temperatures at zero and 
finite chemical potentials.  Note that $d_0 (= T_{pc}/T_{pc}(0)$ at $\mu=0$) 
deviates from one with 1\% because of the disagreement of $\beta_{pc}(0)$ from Ref.~\cite{Ejiri:2009hq}. 
The result is shown in Fig.~\ref{Feb2711fig2} (left panel). 
The line is almost proportional to the quadratic function at small chemical potentials, and 
shows the rapid increase near $\mu_I=\pi/12$. 
We obtain the location of  the RW endpoint $\beta= 1.927(5)$, which corresponds to 
$T/T_{pc}\sim 1.15$. Except for the quadratic function, other three functions are consistent with 
obtained data. The quartic function suffers from large errors. 
The difference between two Pad\'e approximations is found near the RW endpoint.

Obtained pseudo-critical line is analytically continued to $\mu^2>0$ region 
using $\mu_I^2 = -\mu^2$. Th results are shown in the right panel of 
Fig.~\ref{Feb2711fig2}. The curvature at $\hat{\mu}/T_{pc}=0$ of a power series 
of $(\hat{\mu}/\pi T_{pc})^2$ is $t_2 = \pi^2 d_2 = 0.38(12)$. 
The present results are slightly smaller than other studies, see e.g. 
Ref.~\cite{Philipsen:2008gf}

\section{Summary and outlook}

We have investigated the two-flavor QCD phase diagram using the lattice QCD simulation 
with the imaginary chemical potential approach. 
The simulation was performed on the imaginary chemical potential region. 
The properties of the imaginary chemical potential region of the QCD phase diagram  
was discussed. We have derived the pseudocritical line, and the results are analytically 
continued to the real chemical potential region. 

The present calculation was performed with the intermediate quark mass and 
small lattice. The finite volume scaling analysis 
and quark mass-dependence analysis are necessary to confirm the present results. 
In particular, the order of the RW endpoint depends on the 
mass of the quark. The improvement on these points should be done in a future study. 
The finite volume scaling analysis is under progress. 

\section*{Acknowledgment}
We thank P. Cea, L. Cosmai, M. D'Elia, Ph. de Forcrand,  
H. Kouno, S. Motoki, Y. Nakagawa, A. Papa,  T. Saito,
Y. Sakai, T. Sasaki and M. Yahiro  for discussions and comments.
The simulation was performed on NEC SX-8R at RCNP, and NEC SX-9 at CMC, 
Osaka University, and HITACHI SR11000 and IBM Blue Gene/L at KEK.
This work was supported by Grants-in-Aid for Scientific Research 20340055 and 20105003.


\end{document}